# Design of eco-friendly fabric softeners: structure, rheology and interaction with cellulose nanocrystals

E.K. Oikonomou*,[1], N. Christov[2], G. Cristobal[2], C. Bourgaux[4], L. Heux[5], I. Boucenna[1] and J.-F. Berret*[1]

[1]Laboratoire Matière et Systèmes Complexes, UMR 7057 CNRS Université Denis Diderot Paris-VII, Bâtiment Condorcet, 10 rue Alice Domon et Léonie Duquet, 75205 Paris, France
[2]Solvay Research & Innovation Center Singapore, 1 Biopolis Drive, Amnios, Singapore 138622
[4]Institut Galien Paris-Sud - UMR CNRS 8612, Faculté de Pharmacie, Université Paris-Sud XI, 92296 Châtenay-Malabry Cedex, France
[5]Centre de recherches sur les macromolécules végétales, BP 53, 38041 Grenoble cedex 9, France

**Abstract:**
*Hypothesis:* Concentrated fabric softeners are water-based formulations containing around 10 – 15 wt. % of double tailed esterquat surfactants primarily synthesized from palm oil. In recent patents, it was shown that a significant part of the surfactant contained in today's formulations can be reduced by *circa* 50 % and replaced by natural guar polymers without detrimental effects on the deposition and softening performances. We presently study the structure and rheology of these softener formulations and identify the mechanisms at the origin of these effects.

*Experiments*: The polymer additives used are guar gum polysaccharides, one cationic and one modified through addition of hydroxypropyl groups. Formulations with and without guar polymers are investigated using optical and cryo-transmission electron microscopy, small-angle light and X-ray scattering and finally rheology. Similar techniques are applied to study the phase behavior of softener and cellulose nanocrystals considered here as a model for cotton.

*Findings*: The esterquat surfactants are shown to assemble into micron-sized vesicles in the dilute and concentrated regimes. In the former, guar addition in small amounts does not impair the vesicular structure and stability. In the concentrated regime, cationic guars induce a local crowding associated to depletion interactions and leads to the formation of a local lamellar order. In rheology, adjusting the polymer concentration at 1/10th that of the surfactant is sufficient to offset the decrease of the elastic property associated with the surfactant reduction. In conclusion, we have shown that through an appropriate choice of natural additives it is possible to lower the concentration of surfactants in fabric conditioners by about half, a result that could represent a significant breakthrough in current home care formulations.

Corresponding authors: evdokia.oikonomou@univ-paris-diderot.fr, jean-francois.berret@univ-paris-diderot.fr

# I – Introduction

Concentrated fabric softeners are water-based formulations containing around 10 – 15 wt. % of double tailed surfactants, primarily synthesized from palm oil derivatives. These laundry products are used since the 1960's to make cotton fabrics soft, fresh smelling and wrinkle free. For the last decade, efforts have been made to develop conditioners based on natural and less aquatoxic materials [1]. The first step towards more eco-friendly formulations has been taken with the replacement of dialkyldimethylammonium surfactants by esterquats. Esterquats are quaternary ammonium compounds having two (C(16)-C(18)) fatty acid chains with two weak ester linkages. The inclusion of ester linkages has significantly improved the biodegradation kinetics of these materials, lowering the environmental exposure [2]. The second step for the development of eco-friendly products is modern and concerns the reduction of the surfactant concentration used in house and personal care formulations. In this work we explore this possibility to lessen the surfactant content in fabric softeners while preserving the initial deposition and softening performances.

Although softeners were used for decades [3], there is still a lack of consensus pertaining to the active deposition on cotton fibers as well as on the softening mechanism. It has been shown that the surfactants used in conditioners are self-assembled into micron size vesicles [4-6]. In general it is assumed that the softening effect is due to the formation of a lubricating surfactant layer on the fibers [7-9]. Recent studies have correlated fabric frictional characteristics with smoothness and softness [10], while others have put forward that the friction between the human skin and fabrics has a more prominent role [11]. Following Igarashi *et al*. the softening effect would come from the reduction of H-bonding between the bound water molecules and the cotton fibers [12,13]. Crutzen has suggested that hydrophobic interactions derived from the long alkyl chains of cationic surfactants are the primary driving force for adsorption [14], whereas Kumar *et al*. proposed that the deposition process is electrostatically driven and that intact vesicles adsorb on cotton or viscose fibers [5,15]. The deposition of vesicular structures (including uni/multilamellar and mutivesicular vesicles) on solid substrates results in the formation of a so-called supported vesicular layer (SVLs) [8,16]. Understanding the SVL formation is also important for conditioner and active delivery applications. In fabric softeners, surfactant vesicles act as vehicles for driving water insoluble molecules (e.g. fragrances) on cotton [17]. In a recent paper, we have shown that in contact with cellulose nanocrystals, vesicles maintain their shape, in agreement with the work by Kumar an coworkers [5] and are strongly associated to the surfactant membrane *via* electrostatic forces [6]. To produce the next conditioner generation, it is essential to understand the intrinsic processes pertaining to the deposition and softening mechanism, as well as being able to control them through a solid physicochemistry approach.

Recently the Solvay application laboratories have filed a series of patents that describe the benefits of adding minute amounts of polysaccharides to actual formulations in order to improve their softening and fragrance delivery performances [18,19]. Natural hydrocolloid polymers such as cationic guar gum (C-Guar) and a hydroxypropyl guar (HP-Guar) extracted from the seeds of cyamopsis tetragonalobus plant were considered as additives for this application. In these patents, it is reported that towels treated with different formulations, including additive-free benchmark were evaluated from independent panellists in double-blinded tests. The softness of the treated towels was rated in a scale of 1 to 5, wherein 1 represents the lowest softness and 5 represents the highest. According to this survey, esterquat formulations modified with the C-Guar and HP-Guar polymers received the highest ranking for the softening performance, 4.4 against 4.0 for the additive-free benchmark. At the same time the surfactant concentration could be reduced by about half, from 10.5 wt. % to 4 – 6 wt. %.

In this work, we investigate the effects of cationic and hydroxypropyl guar polymers focusing on the structural and rheological properties of novel conditioner formulations. The dispersions are investigated using a combination of techniques including optical and cryo-transmission electron microscopy, small-angle X-ray scattering and rheology. It is found that at low concentration, the polymers do not alter the local structure of the cationic esterquat vesicles, whereas at higher concentrations the phase and rheological behaviors are modified. By adjusting the polymer concentration at around 1/10 that of the surfactant, it is possible to offset the decrease of the elastic property associated with the surfactant reduction. The deposition on model cotton fibers is simulated following the interaction of vesicles and polymers with cellulose nanocrystals [20-23]. Cellulose nanocrystals are here used as a stand-in for cotton to facilitate the assessment of the interactions in bulk phases. We finally show that the physicochemical study of conditioners in the dilute and concentrated regimes, combined with the testing of their deposition and softening performances are useful to help guide the development of novel and environmentally friendly softeners.

# II – Experimental Section
## II.1 – Materials and sample preparation
**Materials.** The esterquat surfactant ethanaminium, 2-hydroxyN,N-bis(2-hydroxyethyl)-N-methyl-esters with saturated and unsaturated C16-18 aliphatic chains, abbreviated TEQ in the following was provided by Solvay®. For TEQ, the gel-to-fluid transition related to the long-range order of the surfactant molecules in the membrane is found at $T_M = 60$ °C [6]. The counterions associated to the quaternized amines are methyl sulfate anions. The polysaccharide polymers are a cationic guar gum (C-Guar, molecular weight $2 \times 10^6$ g mol$^{-1}$) and a hydroxypropyl guar gum (HP-Guar, molecular weight $0.5 \times 10^6$ g mol$^{-1}$), both synthesized by Solvay® (Scheme 1). Guar is a natural polymer extracted from the seeds of cyamopsis tetragonalobus plant. HP-Guar is widely used in oil and gas recovery, paints, cosmetics and biology [24] as a thickening and lubricant agent [25,26]. Cationic guar is known for its conditioning properties, notably on hair [27-29]. They were obtained by introducing positively charged trimethylamino(2-hydroxyl)propyl into the backbone.

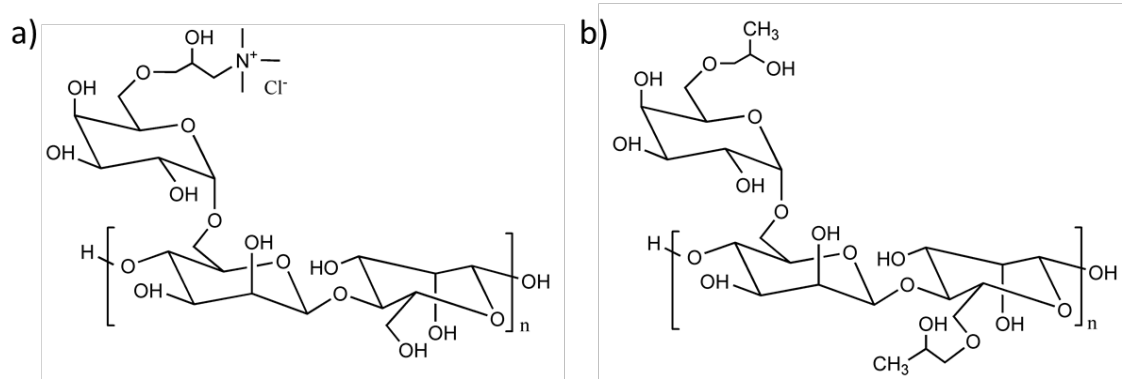

**Sheme I:** *Chemical structures of the cationic (a) and hydroxypropyl modified (b) guar polymers used in this work. These polymers are abbreviated as C-Guar and HP-Guar respectively.*

Cellulose nanocrystals (CNC) was prepared according to earlier reports using catalytic and selective oxidation.[30] Briefly, cotton linters provided by Buckeye Cellulose Corporation were hydrolyzed according to the method described by Revol *et al.* treating the cellulosic substrate with 65% (w/v) sulfuric acid at 63 °C during 30 min [30]. The suspensions were washed by repeated centrifugations, dialyzed against distilled water until neutrality and sonicated for 4 min with a Branson B-12 sonifier equipped with a 3 mm microtip. The suspensions were then filtered through 8 μm and then 1 μm cellulose nitrate membranes (Whatman). At the end of the process, a 2 wt. % aqueous stock suspensions was obtained. The resulting nanoparticles are in the form of laths and have a length of 180 ± 30 nm, a width of 17 ± 4 nm and a height of 7 nm. These values were confirmed by Cryo-TEM **(see Supplementary Fig. S1)**. Using light scattering, the CNC dispersion prepared at pH 4.5 and at a concentration of 0.1 wt. % displays a single relaxation mode in the autocorrelation function. $g^{(2)}(t)$ is decreased rapidly above delay times around $10^3$ μs, indicating that the nanofibers are not aggregated while the size studies revealed an intensity distribution centered around 120 nm **(see Supplementary Fig. S1)**. Water was deionized with a Millipore Milli-Q Water system. All the products were used without further purification.

**Formulation preparation**. All concentrated aqueous formulations were prepared using MilliQ filtered water at 60 °C following the protocol described in Refs. [18,19]. TEQ was first melted at 60 °C and was added dropwise in water. The final pH was adjusted at 4.5. The dispersions containing guar polymers were produced by first adding the guars in 60 °C water and then the melted TEQ. The pH was adjusted at 4.5 after the guar addition and at the final step. Five formulations were prepared: TEQ 10.5 wt. %, TEQ 6 wt. %, TEQ 6 wt. % + C-Guar 0.4 wt. %, TEQ 6 wt. % + HP-Guar 0.4 wt. % and TEQ 6 wt. % + C-Guar 0.3 wt. % + HP-Guar 0.3 wt. %. Also, for SAXS studies TEP : HP-Guar : C-Guar = 4 : 0.3 : 0.3 were prepared. The investigated dispersions were produced by dilution down to the desired TEQ concentration using MilliQ filtered water ($T$ = 25 °C). Finally, for polymer characterization, aqueous dispersions of each guar (pH 4.5) were prepared by diluting a stock aqueous solution of 1 wt. % prepared at 60 °C water. Unfiltered and 0.45 μm filtered (cellulose acetate) diluted dispersions were used for characterization.

**Mixing protocol.** The interactions between CNC and guars or CNC and TEQ/guar mixtures were investigated using the direct mixing formulation pathway [21,31,32]. Batches of CNC, guars or TEQ/C-Guar (dilution from TEQ 6 wt. % + C-Guar 0.4 wt. % at TEQ 0.01 wt. %) were prepared in the same conditions of pH (pH 4.5) and concentration ($c$ = 0.01 and 0.1 wt. %) and then mixed at different ratios, noted $X = V_{Soft}/V_{CNC}$, where $V_{Soft}$ and $V_{CNC}$ denote the volumes of the softener formulation (either guar or TEQ + guar) and CNC dispersion, respectively. The mixed dispersions were prepared at room temperature (25 °C). After mixing, the dispersions were stirred rapidly, let to equilibrate for 5 minutes and the scattered intensity and hydrodynamic diameter were measured in triplicate. As the concentrations of the stock solutions are identical, the volumetric ratio $X$ is equivalent to the mass ratio between constituents. In the figures, the concentration of the stock solutions are set at $X = 10^{-3}$ and $X = 10^3$.

## II.2 – Experimental techniques

**Light scattering.** The scattering intensity $I_S$ and hydrodynamic diameter $D_H$ were measured using the Zetasizer Nano ZS spectrometer (Malvern Instruments, Worcestershore, UK). A 4 mW He−Ne laser beam ($\lambda$ = 633 nm) is used to illuminate the sample dispersion, and the scattered intensity is collected at a scattering angle of 173°. The Rayleigh ratio $\mathcal{R}$ was derived from the intensity according to the relationship: $\mathcal{R} = (I_S - I_w)n_0^2 \mathcal{R}_T / I_T n_T^2$ where $I_w$ and $I_T$ are the water and toluene scattering intensities respectively, $n_0$ = 1.333 and $n_T$ = 1.497 the solution and toluene refractive indexes, and $\mathcal{R}_T$ the toluene Rayleigh ratio at $\lambda = 633$ nm ($\mathcal{R}_T = 1.352 \times 10^{-5}\ cm^{-1}$) [33]. The second-order autocorrelation function $g^{(2)}(t)$ is analyzed using the cumulant and CONTIN algorithms to determine the average diffusion coefficient $D_C$ of the scatterers. The hydrodynamic diameter is then calculated according to the Stokes-Einstein relation, $D_H = k_B T / 3\pi\eta D_C$, where $k_B$ is the Boltzmann constant, $T$ the temperature and $\eta$ the solvent viscosity. Measurements were performed in triplicate at 25 °C after an equilibration time of 120 s.

**Zeta Potential**. Laser Doppler velocimetry (Zetasizer, Malvern Instruments, Worcestershore, UK) using the phase analysis light scattering mode and detection at an angle of 16° was performed to determine the electrophoretic mobility and zeta potential of the different dispersions studied. Measurements were performed in triplicate at 25 °C, after 120 s of thermal equilibration.

**Single angle X-ray scattering.** Samples were loaded into quartz capillaries (diameter 1.5 mm, Glass Müller, Berlin, Germany). The top of the capillaries was sealed with a drop of paraffin to prevent water evaporation. The temperature of the sample holder was set to 25°C. The scattered intensity was reported as a function of the scattering vector $q = 4\pi sin\theta/\lambda$, where $2\theta$ is the scattering angle and $\lambda$ the wavelength of the incident beam. Small-Angle X-ray Scattering (SAXS) experiments were performed on the SWING beamline at the SOLEIL synchrotron source (Saint-Aubin, France). SAXS patterns were recorded using a two-dimensional AVIEX CCD detector placed in a vacuum detection tunnel. The beamline energy was set at 12 keV and the sample-to-detector distance was fixed to cover the $0.08 - 8$ nm$^{-1}$ $q$-range. The calibration was carried out with silver behenate. The acquisition time of each pattern was 250 ms; 5 acquisitions were averaged for each sample. Intensity values were normalized to account for beam intensity, acquisition time and sample transmission. Each scattering pattern was then integrated circularly to yield the intensity as a function of the wave-vector. The scattered intensity from a capillary filled with water was subtracted from the sample scattering curves.

**Optical Microscopy**. Phase-contrast images were acquired on an IX73 inverted microscope (Olympus) equipped with 20×, 40×, and 60× objectives. Seven microliters of TEQ dispersion were deposited on a glass plate and sealed into a Gene Frame (Abgene/ Advanced Biotech) dual adhesive system. An Exi-Blue camera (QImaging) and Metaview software (Universal Imaging Inc.) were used as the acquisition system.

**Fluorescent Microscopy**. For fluorescent microscopy, 100 µl of TEQ 0.1 wt. % dispersion were mixed with 100 µl of 1 µM PKH67 (Sigma) solution. PKH67 is used in cellular biology as a green fluorescent molecular linker developed for cell membrane labeling. It is characterized by a absorption maximum at 490 nm and an excitation maximum at 502 nm. The solution was then stirred and let equilibrate for 1 h before use. The fluorescent vesicles were then mixed with C-Guar and CNC at the adequate mixing ratios.

**Cryogenic transmission electron microscopy (Cryo-TEM)**. A few microliters of the samples were deposited on a lacey carbon coated 200 mesh (Ted Pella). The drop was blotted with a filter paper on a Vitrobot™ (FEI) and the grid was quenched rapidly in liquid ethane, cooled with liquid nitrogen, to avoid the crystallization of the aqueous phase. The membrane was finally transferred into the vacuum column of a TEM microscope (JEOL 1400 operating at 120 kV) where it was maintained at liquid nitrogen temperature thanks to a cryo holder (Gatan). The magnification was selected between 3000× and 40000×, and images were recorded with a 2k-2k Ultrascan camera (Gatan).

**Rheology**. The rheological studies were performed with a Physica RheoCompass MCR 302 (Anton Paar) using a cone-and-plate geometry (diameter 50 mm, cone angle 1°) and a solvent trap to prevent water evaporation. The linear viscoelastic regime was determined by performing a strain sweep test at a frequency of 1 rad s$^{-1}$. The frequency dependent storage and loss moduli, $G'(\omega)$ and $G''(\omega)$, were determined by applying an oscillatory shear stress of 1 rad s$^{-1}$.

# III - Results and Discussion
## III.1 - Partial replacement of TEQ surfactant with natural guar polymers
### III.1.1 - Guar polymer characterization

Here we investigate the guar polymers used in the patented formulation described in the Introduction section [18]. Two different guar polymers were used in this study, a cationic (C-Guar) and a hydroxypropyl (HP-Guar) guar. The C-Guar is believed to enhance the deposition on fabrics thanks to its positive charges, while HP-Guar was chosen for its rheological properties. For the characterization, aqueous guar dispersions were prepared at different concentrations (0.001 – 1 wt. %) and were studied by light scattering and $\zeta$-potential. The second-order autocorrelation functions $g^{(2)}(t)$ of filtered C-Guar and HP-Guar dispersions at 0.02 wt. % are shown as a function of delay time in Figs. 1a and 1b respectively. The intensity distributions in the insets reveal a single peak around 200 nm for C-Guar and a double peak for HP-Guar at 50 nm and 350 nm. These hydrodynamic diameters are larger than those expected from polymers of molecular weight $M_W$ = 0.5 and 2×10$^6$ g mol$^{-1}$ respectively [34], suggesting that the chains may by associated in water and form hydrocolloid particles [17]. Electrophoretic mobility measurements on dilute dispersions give zeta potential of +30 mV and 0 mV respectively.

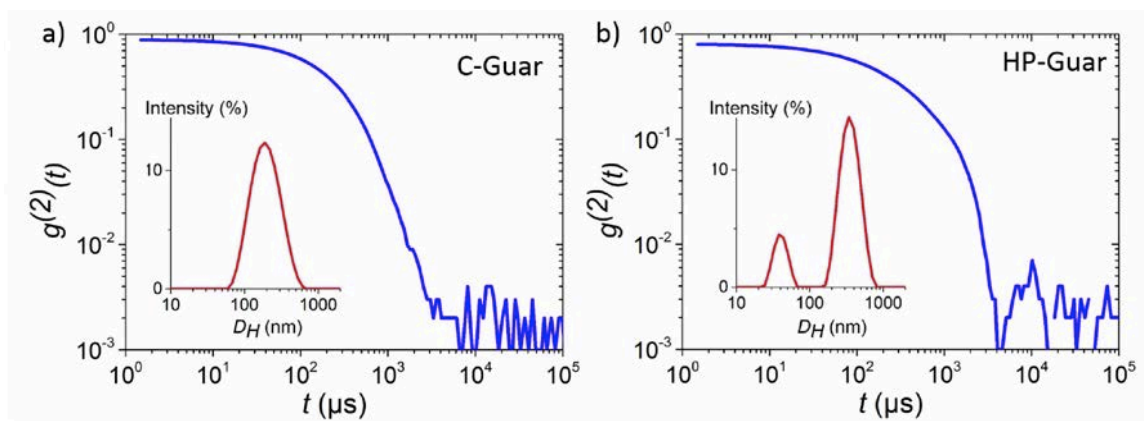

*Figure 1*: Second-order auto-correlation functions and size distribution of a) C-Guar and b) HP-Guar aqueous filtered dispersions at 0.02 wt. %.

### III.1.2 - Effect of guars on the vesicle structure at low concentration
Mixed surfactant/guar dispersions were first studied by phase-contrast optical microscopy in the diluted regime. The dispersions were obtained by diluting 6 wt. % TEQ formulations with DI-water at pH 4.5. For the sample containing the polysaccharides, the C-Guar and HP-Guar concentrations were 0.3 wt. %. Fig. 2a and 2b show optical microscopy images of diluted TEQ dispersions in the conditions of use ($c_{TEQ}$ = 0.01 wt. %) without and with guar polymers, respectively. Both figures exhibit well-contrasted spherical particles of average size 800 nm in both cases (Fig. 2b and 2d). These objects were identified as TEQ vesicles (see **Supplementary Movies 1 and 2** illustrating the Brownian motion of the vesicles). This finding shows that at the micron scale there is no apparent difference between vesicles with or without guar polymer and that TEQ vesicles are stable upon guar addition.

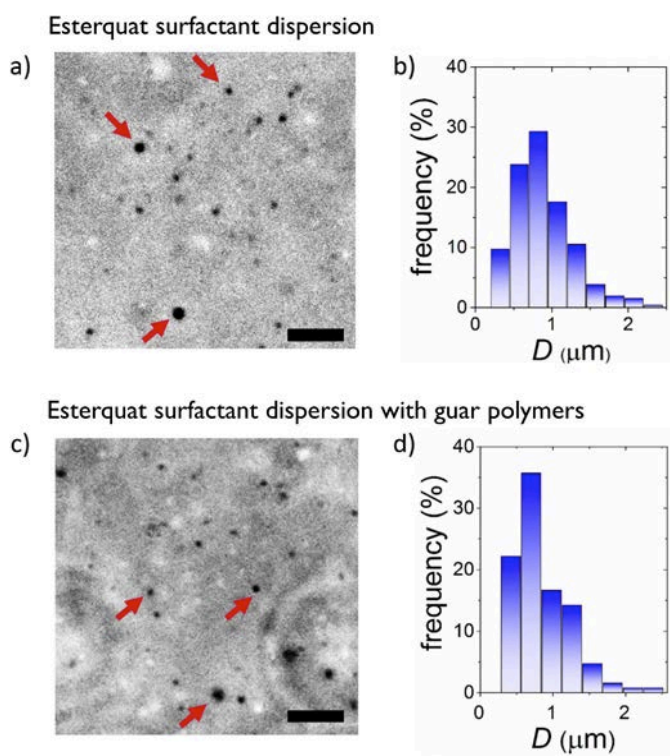

*Figure 2*: *a) Phase-contrast microscopy image of a 0.01 wt. % TEQ dispersion and b) associated vesicular size distribution. c and d), same as a) and b) for a 0.01 wt. % TEQ + guars dispersion. The dispersions without polymers was obtained by diluting 6 wt. % TEQ formulation with DI-water at pH 4.5. The dispersion containing the polysaccharides was prepared at $c_{TEQ}$ = 6 wt. % with 0.3 wt. % for each guar. Scale bars are 10 µm.*

These observations were further confirmed by cryo-TEM. Representative images from the dilute regime (0.1 wt. %) for both formulations are shown in Fig. 3a and 3b. It is found that the TEQ surfactants self-assemble locally into a bilayer structure. The bilayers close on themselves to form unilamellar, doublets and multivesicular vesicles of size in the range of 100 – 300 nm. Multivesicular vesicles are defined as a membrane compartment that encapsulates several smaller vesicles [35]. It should be noted that in contrast to earlier studies [4,36] multi-lamellar vesicles in the form of "onions" were not observed for this surfactant **(see Supplementary Fig. S2)**. Electrophoretic mobility measurements show that the TEQ vesicles are positively charge, with a zeta potential of $\zeta$ = + 65 mV. The average size and zeta potential data determined for the different species studied in this work, including the C-Guar and HP-Guar polymers, the TEQ vesicles and the cellulose nanocrystals are summarized in Table I. In conclusion, cryo-TEM reveals similar structures for the two dilute formulations and confirms that the TEQ-bilayers with and without polymers are stable upon dilution in the conditions of use.

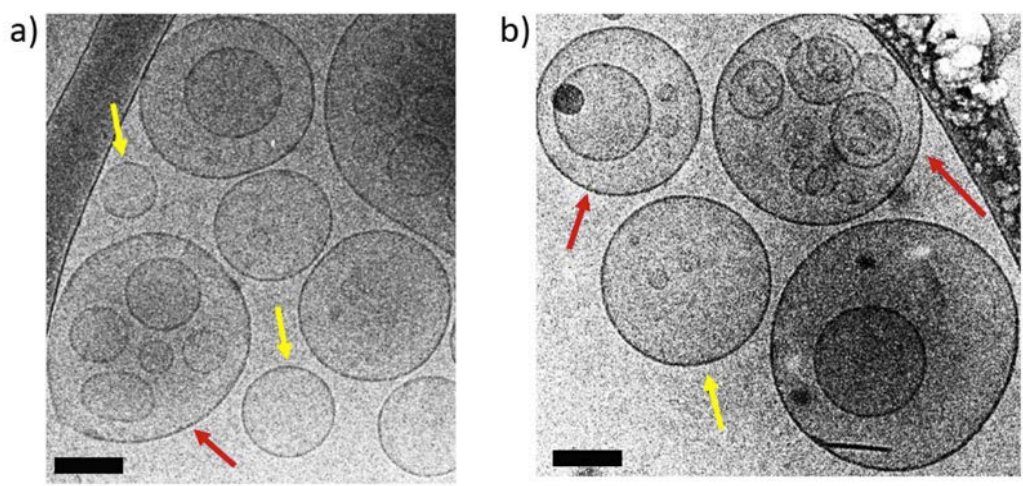

*Figure 3:* Representative cryogenic transmission electron microscopy (cryo-TEM) images of 0.1 wt. % TEQ surfactant aqueous dispersions a) without guar polymers and b) with C-Guar and HP-Guar at 0.01 wt. %. Yellow and red arrows indicate the unilamellar and multi-vesicular vesicles respectively. Scale bars are 100 nm.

|  | Characteristic size | Zeta potential ζ |
|---|---|---|
| Cationic guar polymer (C-Guar) | 200 nm | + 30 mV |
| HP-Guar polymers | 50 nm / 350 nm | 0 mV |
| TEQ surfactant vesicles | 100 nm – 10 μm | + 65 mV |
| Cellulose nanocristal (CNC) | 180 nm* | -38 mV |

*Table I:* List of the characteristic features obtained for the polymers, vesicles and celluloses used in this study. The characteristic size was determined from light scattering, optical microscopy and transmission electron microscopy. *The cellulose nanocrystals are anisotropic in the form of laths with dimensions $180 \times 17 \times 7$ $nm^3$.

### III.1.3 - Effect of guars on the vesicle structure in formulations in the concentrated regime

The effect of polymer guars at the formulation concentration is first examined by optical microscopy. Fig. 4 displays a series of 40× magnification images of TEQ dispersions at concentration 10.5 wt. % (Fig. 4a) and 6 wt. %, this later being without and with guar polymers (Figs. 4c and 4d respectively). The TEQ 10.5 wt. % dispersion exhibits highly contrasted and densely packed vesicles characterized by broad dispersity, with sizes comprised between 200 nm to 10 μm (Fig. 4b). Some large vesicles are also not spherical and exhibit facets (Fig. 4a), a phenomenon attributed to mechanical stresses generated during preparation and related with the liquid-crystalline order of the alky chains [37-39]. In **Supplementary Movie 3**, the large vesicles appear to be immobile and stuck to the glass surface. Numerous smaller vesicles are visible however and animated of rapid Brownian motion. At 6 wt. %, the TEQ vesicles are less packed and exhibit again Brownian motion **(see Supplementary Movie 4)**. The appearance of the guar/surfactant mixed dispersions in Fig. 4d is different: neither large faceted nor single vesicles are observed and the previous patterns are replaced by a heterogeneous micron sized texture. This texture appears moreover to be frozen at the time scale of the observations (> minutes, **see Supplementary Movie 5**). To get a better insight into these dispersions, SAXS and rheology measurements were performed.

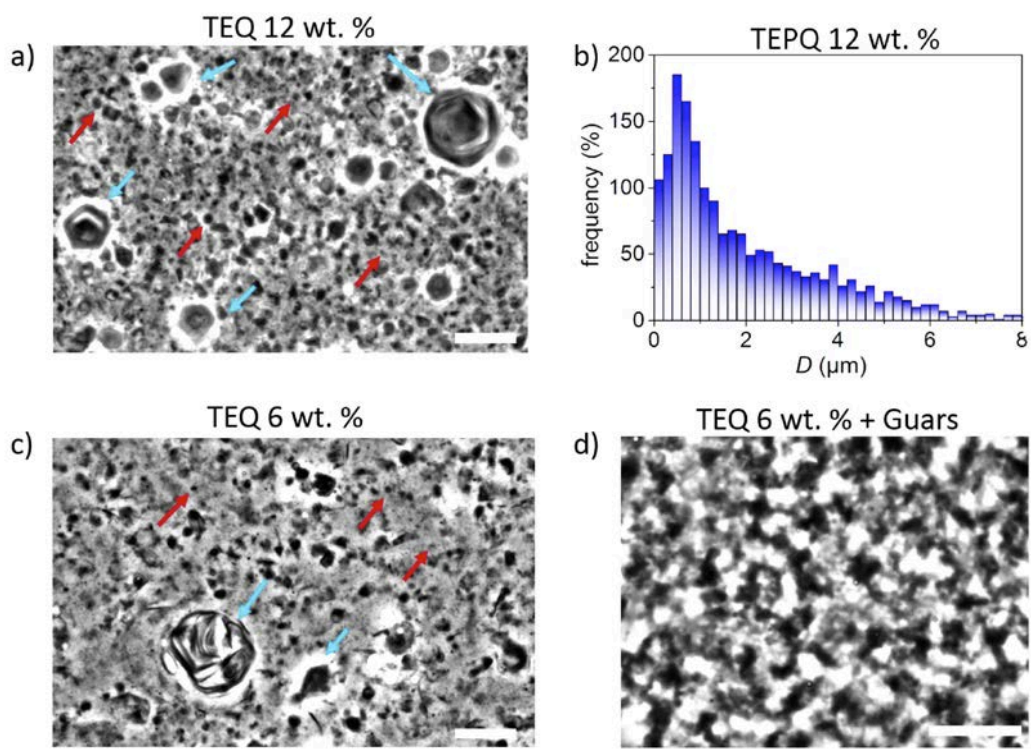

*Figure 4*: TEQ surfactant dispersions studied by phase-contrast microscopy: a) image at 10.5 wt. % and b) the corresponding vesicular size distribution, c and d) microscopy images at TEQ 6 wt. % without and with 0.6 wt. % of guar polymers (0.3 wt. % of C-Guar and 0.3 wt. % of HP-Guar). Scale bars are 10 µm. Blue arrows: large faceted vesicles. Red arrows: micrometric size vesicles.

Figs. 5a, 5b and 5c display the SAXS patterns for 4 wt. % TEQ aqueous dispersions without and with polymers respectively, the guar concentrations being 0.2 wt. % and 0.6 wt. % for each. Data obtained from pure polymer dispersions (0.6 wt. %) are also included for comparison. The large oscillation in the $q$-range $0.4 - 2.5$ nm$^{-1}$ observed in Fig. 5a is characteristic of the TEQ bilayer form factor and arises from the electronic density modulation across the bilayer [40,41]. The bilayer form factor is modeled using a two-level electronic density profile (inset), one for the head groups (0.8 nm) and the other for the alkyl chains (3.0 nm). As a result, the overall layer thickness is estimated at 4.6 nm, in good agreement with the cryo-TEM determination [6]. The continuous line labeled *Model* in Fig. 5a proves that this approach is well founded. For the polymers, the scattering cross-section exhibits a decrease consistent with that of polymer

material, in particular the observation of $q^{-2}$ power law at large wave-vector [34]. The range studied does not allow however to determine the polymer gyration radius.

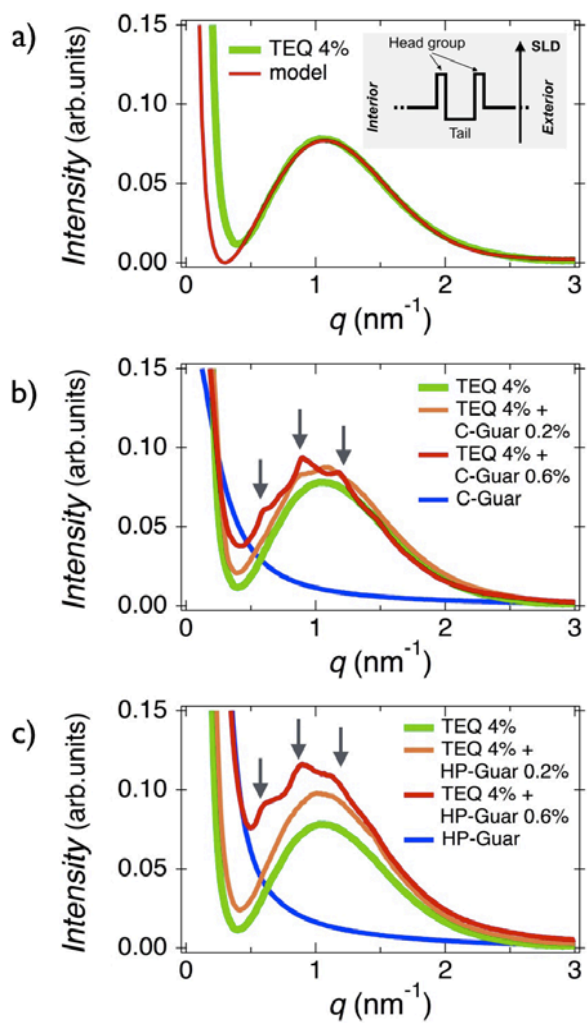

*Figure 5: Small-angle X-ray scattering from TEQ 4 wt. % (a) and TEQ 4 wt. % in presence of 0.2 wt. % or 0.6 wt. % guar polymers (b, c). In a) the bilayer form factor is adjusted using a 4.6 nm membrane with a two-level scattering length density (SLD) displayed in the inset. The spectra in b) and c) show the effect upon C-Guar and HP-Guar addition respectively. The arrows indicate the values of wave-vectors where a Bragg peak is present.*

Upon addition of 0.2 wt. % of guar, the TEQ form factor is not altered, suggesting no significant change in the TEQ bilayer. At 0.6 wt. % however, low amplitude Bragg peaks superimposed to the bilayer form factor are observed [40]. For TEQ + C-Guar these peaks are located at $q_2 = 0.6$, $q_3 = 0.9$ and $q_4 = 1.18$ nm$^{-1}$ and at $q_2 = 0.62$, $q_3 = 0.91$ and $q_4 = 1.14$ nm$^{-1}$ for TEQ + HP-Guar. These peaks are reminiscent of the maxima observed in TEQ concentrated dispersions at 8, 12 and 15 wt. %. In our previous study, these peaks were interpreted in terms of a structure factor arising from stacked bilayers [6]. Note here that the first diffraction order at $q_1$ is absent, probably because it is down modulated by the bilayer form factor oscillation at 0.4 nm$^{-1}$. In terms of inter-lamellar distances, one finds 20.9 nm for TEQ + C-Guar and 20.3 nm for TEQ + HP-Guar. Interestingly, the positions of the Bragg peaks found here are similar to those observed in pure TEQ dispersion at 8 wt. %. So, adding 0.6 wt. % of guar to a 4 wt. % TEQ dispersion or doubling the TEQ concentration has a similar effect on the vesicular structure. In summary we have found that for concentrated dispersions the guars have a significant impact on the vesicle microstructure. The presence of Bragg peaks suggests a transformation from multi-vesicular to multilamellar vesicles upon polymer addition. We surmise that crowding effects are triggered by the increase in overall volume fraction and/or by polymer induced depletion interactions [38,39]. Adding polymers to a colloidal dispersion is known to induce short-range attractive forces between particles above a threshold volume fraction [42,43]. It is also possible that other interaction types are present and also modify the vesicular organization.

*III.1.4 - Effect of guars on the surfactant formulation rheology*

The elastic and loss moduli $G'(\omega)$ and $G''(\omega)$ were determined experimentally for TEQ surfactant solutions at 10.5 wt. % (Fig. 6a) and 6 wt. % (Fig. 6b). It is observed that the concentrated TEQ solution exhibits gel-like behavior, as $G'(\omega) > G''(\omega)$. This structural arrangement imparts to the fluid the rheological properties of a soft solid characterized by a gel-like elastic complex modulus and a non-zero yield stress. From the stress *versus* rate flow curve, the yield stress was estimated at 0.1 Pa [6], that is typically 40 times lower than the elastic modulus at 1 rad s$^{-1}$ **(see Supplementary information S3)**. In contrast, the 6 wt. % TEQ solution exhibits a liquid-like behavior (*i.e.* $G''(\omega) > G'(\omega)$) with values for storage and loss moduli around 0.01 Pa. These features indicate that upon dilution the formulation has lost its soft solid character. These results are in agreement with the optical microscopy where for TEQ 6 wt. % vesicles undergo rapid Brownian motion **(see movie#2)**. Upon guar addition, the rheological behavior of the 6 wt. % dispersion changes drastically. As shown in Figs. 6c and 6d, the moduli are both increased and $G'$ is now higher than $G''$. We also notice that the effect is stronger for HP-guar than for C-Guar. With the combination of the two polymers, the rheological properties are further enhanced and similar to the initial 10.5 wt. % formulation (Fig. 6e). At the concentration used (0.4 wt. %), the aqueous guar dispersions alone are viscous polymeric fluids with a slight viscoelastic behavior **(see Supplementary Fig. S4)**. The high viscosity of the guars arises from H-bonding intermolecular association between the hydrocolloids [44]. In conclusion, we interpret the guar viscosifying effects as a combination of two factors. On one side, in the concentrated phase, the guars induce vesicular crowding and/or aggregation driven by depletion interaction. In addition to this structural change, the HP-Guar hydrocolloid network enhances the background viscosity endowing the system with the appropriate rheological properties.

## III. 2. Cellulose nanocrystal/guar interaction

In our previous report, we propose a method to evaluate cellulose/surfactant interactions with increased sensitivity [6]. The method is based on the use of cellulose nanocrystals instead of micron-sized fibers or yarns and of a combination of different bulk characterization techniques. As a result, we could show that cellulose nanocrystals interact strongly with the TEQ vesicles *via* electrostatic charge complexation leading to the formation of large-scale aggregates in which the vesicles remain intact and adsorbed at the cellulose surface [6]. Here we extend this approach and investigate the interaction of CNCs with guar polymers using the continuous variation method [22,23,45,46]. C-Guar/CNC dispersions were prepared by mixing stock solutions at different volumetric ratios between 10$^{-3}$ and 10$^{3}$. In Fig. 7a, the Rayleigh ratio obtained at 0.01 wt. % exhibits a marked maximum centered around $X_{Max}$ = 0.2. The continuous red line is calculated assuming that $\mathcal{R}(X)$ is the sum of the Rayleigh ratios $\mathcal{R}_{CNC}$ and $\mathcal{R}_{C-Guar}$ weighted by their respective concentrations, where $\mathcal{R}_{CNC}$ and $\mathcal{R}_{C-Guar}$ are the Rayleigh ratios of the CNCs and polymer dispersions [20,21]. $\mathcal{R}(X)$ is found to be higher than the prediction for non-interacting species. The occurrence of a maximum suggests that C-Guar/CNC aggregates are characterized by a fixed stoichiometry, likely related to the point of zero charge [33,35]. The scattered intensity is high ($\mathcal{R} > 10^{-4}$ cm$^{-1}$) and indicates a strong interaction behavior. These results were confirmed by the hydrodynamic diameter measurement that also displays a characteristic peak at $X_{Max}$ **(see Supplementary Fig. S5)**.

The distribution intensities *versus* zeta potential obtained from electrophoretic mobility measurements between $X = 0$ and $X = 100$ are presented in Fig. 7b. Well-defined intensity peaks are found to shift progressively from a negative value ($\zeta$ = - 38 mV) for pure CNC to a positive one ($\zeta$ = + 30 mV) for C-Guar solutions (see Table I). The point of zero charge is found at $X = 0.5$, close to the scattering maximum (Fig. 7a) [32,33]. The results suggest an interaction based on electrostatic association between opposite charged species [20,46,47]. There, the cationic guar particles are found to adsorb onto the CNCs, eventually reversing the sign of the electrophoretic mobility when added in excess and inducing aggregation at charge ratio close to unity (Fig. 7d). In contrast, weak interaction was found between CNC and HP-Guar, as exemplified in **Supplementary Fig. S6**.

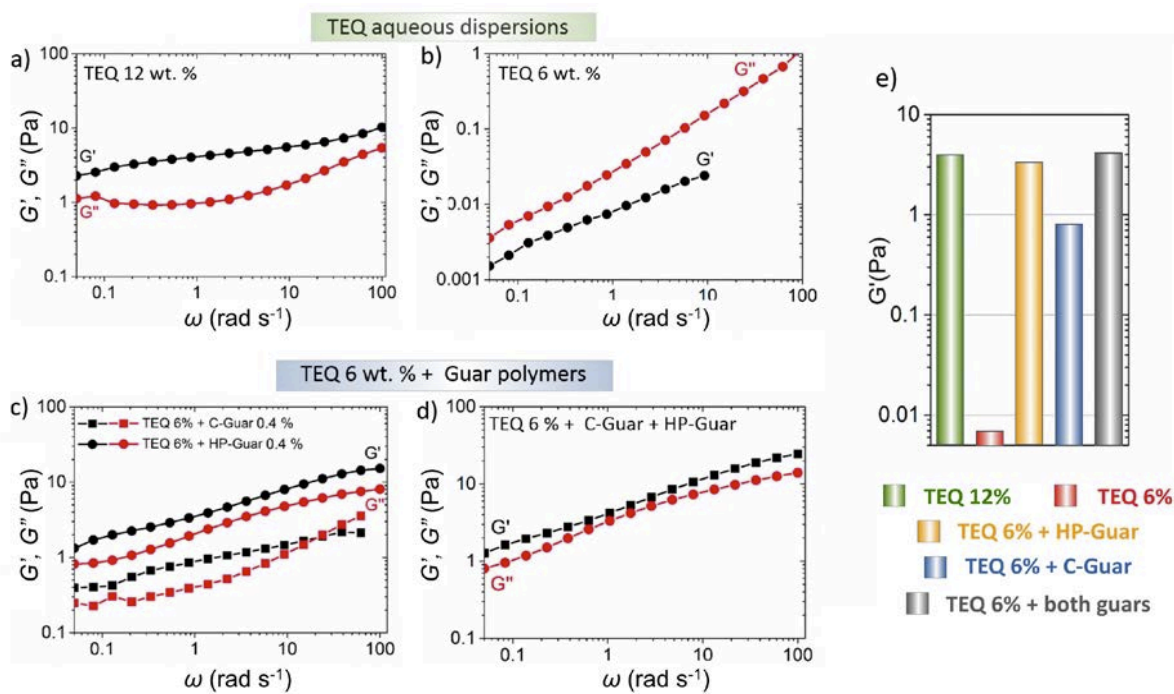

*Figure 6*: *Elastic and loss modulus $G'(\omega)$ and $G''(\omega)$ as a function of the angular frequency for TEQ surfactant at a) 10.5 wt. %, b) 6 wt. %, c) 6 wt. % in presence of C-Guar or HP-Guar 0.4 wt. % and d) 6 wt. % in presence of both C-Guar 0.2 wt.% and HP-Guar 0.2 wt.%. e) $G'(\omega)$ values of the formulations at 1 rad s$^{-1}$.*

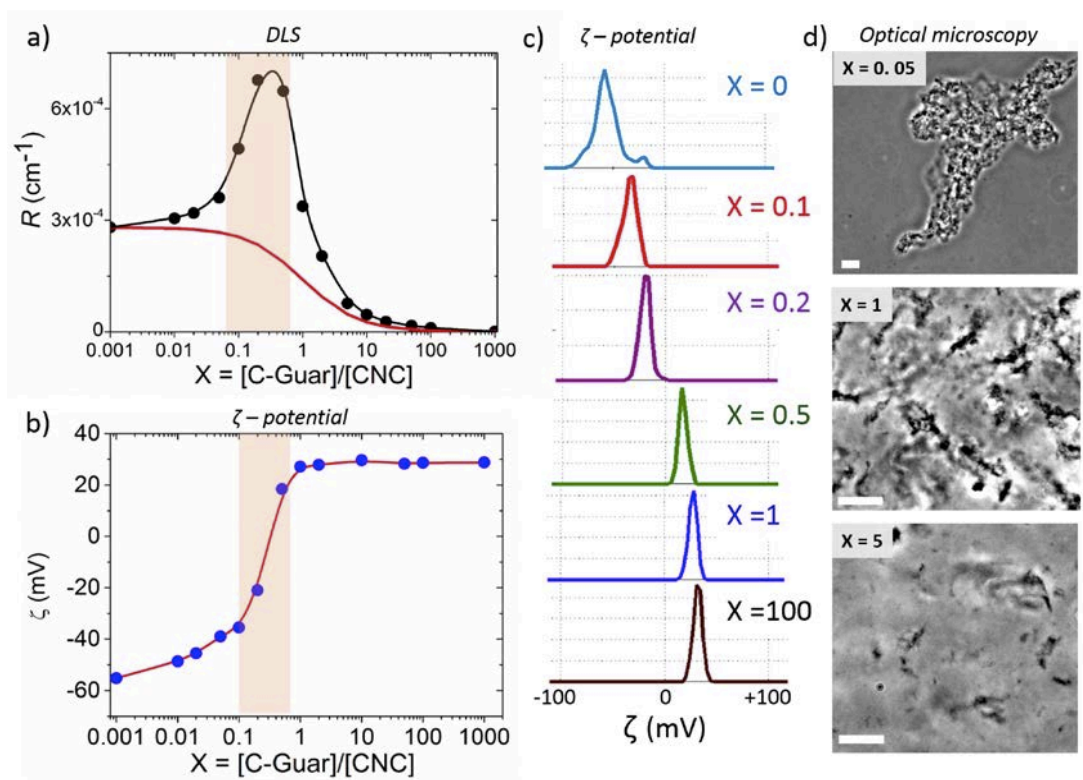

*Figure 7:* a) Rayleigh ratio $\mathcal{R}(X)$ and b) zeta potential results $\zeta(X)$ of mixed C-Guar / cellulose nanocrystal dispersions as a function of X. The continuous line in red in a) is calculated from the non-interacting model (see text). The continuous lines in black and red in a) and b) respectively are guides to the eyes. c) Intensity distributions as a function of the zeta potential $\zeta$ for samples between X = 0 and X = 100 showing a displacement of the peak from negative (pure cellulose nanocrystals) to positive (pure C-Guar) zeta values. c = 0.01 wt. %. d) Optical microscopy images of mixed C-Guar / cellulose nanocrystal dispersions at different mixing ratios X (here c = 0.1 wt. % and the scale bars are 20 µm).

So far, we have evaluated the interaction between the different species in a pairwise manner: surfactant and guar, surfactant and nanocellulose and in the previous section guar and nanocellulose. We now investigate ternary dispersion containing the three components. To this aim, the guar to surfactant ratio is fixed at its actual formulation value, 6.0, 0.30 and 0.30 for TEQ, C-Guar and HP-Guar respectively. The ternary stock solutions were prepared by dilution at a TEQ concentration of 0.01 wt. % and were mixed with a 0.01 wt. % nanocellulose aqueous solution at different volumetric ratios. The results for TEQ + C-Guar mixed with CNC are shown in Fig. 8. It is found that at a ratio of about 0.5 the dispersions precipitate (green shaded area in Fig. 8a). Similar results were found for TEQ/HP-Guar mixed dispersions **(see Supplementary Fig. S7)**. From this result it is concluded that both guars enhance the interaction of TEQ vesicles with nanocellulose. We recall that TEQ/CNC solutions prepared in the same conditions do not display a phase separation [6]. These findings indicate synergistic effects between TEQ and the guars and suggest that in real conditions guar polymers not only adsorb on cotton but also increase the TEQ adsorption. Fluorescent microscopy studies on the mixed aggregates were conducted to evaluate the size and morphology of the separated phases and to test whether the phase separation is of liquid-liquid (coacervation) or of liquid-solid type [32]. TEQ bilayers were labeled with the fluorescent dye PKH67, a lipid molecule used in cellular biology and known to insert into biological membranes. In Figure 8b, a mixture of TEQ+ C-Guar/CNC ($X = 1$, $c = 0.1$ wt. %) is observed by phase contrast and fluorescent optical microscopy. Micron-sized aggregates reveal the coexistence of bright spots identified as TEQ vesicles (arrows) and a green fluorescence continuum and confirm the liquid-solid character of the transition.

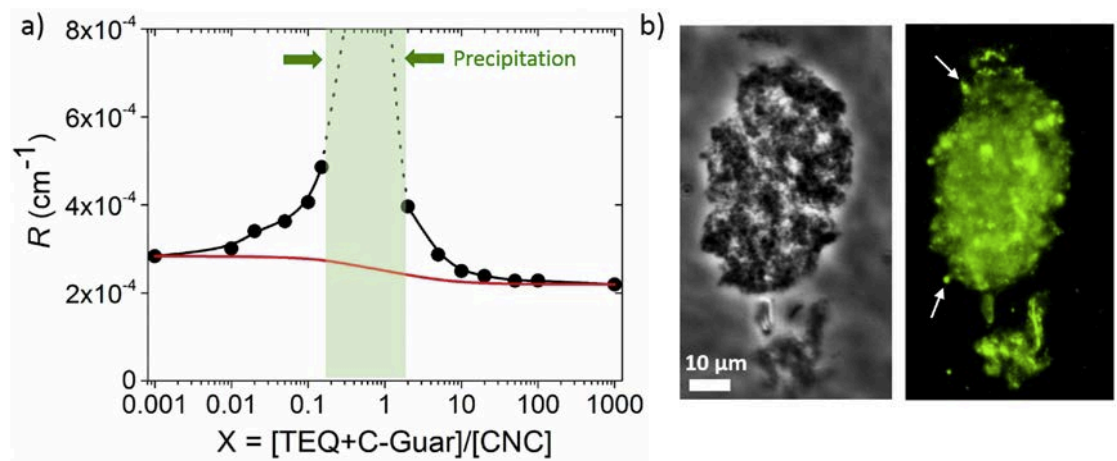

*Figure 8:* a) Rayleigh ratio $\mathcal{R}(X)$ for TEQ+C-Guar / cellulose nanocrystal dispersions (c = 0.01 wt. %) as a function of the mixing ratio X and b) Phase-contrast and fluorescence optical microscopy mixed TEQ+C-Guar / cellulose nanocrystal aggregates (c = 0.1 wt. %). The scale bar is 10 μm.

# IV - Conclusion

On the basis of the previously reported preparative approaches [6,18,19], this work demonstrates that fabric conditioner formulations containing reduced amounts of surfactants, typically from 10-15 wt. % to 4-6 wt. % can be produced while preserving their overall physicochemical properties. To this aim, part of the surfactant is replaced by minute amount of natural guar polymers obtained from the seeds of cyamopsis tetragonalobus plants. In the formulation, the guars are chemically modified with cationic charges or with hydroxypropyl groups, each function having its own specific effect in the dispersion. In previous reports, it was shown that TEQ esterquat dispersions formulated with guar polymers around 4-6 wt. % exhibited excellent softening and fragrance delivery performances, with softness rated from testers at 4.4 on a 1-to-5 scale in double-blinded tests. These results have led to the filing of several patents for this invention [18,19]. Here, we investigate the same dispersions as in the patents focusing on their structural and rheological properties. The guar effects are monitored by a combination of different techniques including optical microscopy, cryo-TEM, SAXS and rheology. The dispersions are studied both in the concentrated and dilute regimes of concentration, mimicking the storage and use conditions of these products. The main result of this study is that at low concentration, *i.e.* around 0.01 wt. % the polymers do not alter the local structure of the cationic esterquat vesicles. At 4-6 wt. % however the polymers modify the phase and rheological behaviors. Concerning the structure, SAXS suggests that the cationic guars induce a local crowding associated to depletion interactions and lead to the formation of a local lamellar order. Adjusting the polymer concentration around 0.2 wt. % for each guar is sufficient to compensate the decrease in storage modulus associated with the surfactant reduction. Recently, we also proposed a method to evaluate cellulose/surfactant interactions with increased sensitivity [6].

This method is applied here to assess the effect of the guar polymers related to the surfactant deposition on cellulose. Interaction studies confirm the high affinity of the surfactant vesicles with the nanofibers. The interaction with the cellulose surface is mainly driven by the oppositely charged surfaces and leads clear phase separations. Concerning the mechanism, it is proposed that in solution the vesicles adsorb spontaneously on cellulose nanofibers and that their structures remain intact. For the next step, further experiments dealing with the deposition and drying of the formulation on actual cotton fabrics should be performed to confirm these assumptions. In conclusion, we have shown that it is possible to reduce the concentration of cationic surfactant present in fabric softeners by half and to preserve the overall physicochemical properties of the formulations, a result that should pave the way for the development of more environmental friendly softeners.

# Acknowledgements

We thank Chloé Puisney, Victor Baldim and Fanny Mousseau for fruitful discussions. ANR (Agence Nationale de la Recherche) and CGI (Commissariat à l'Investissement d'Avenir) are gratefully acknowledged for their financial support of this work through Labex SEAM (Science and Engineering for Advanced Materials and devices) ANR 11 LABX 086, ANR 11 IDEX 05 02. We acknowledge the ImagoSeine facility (Jacques Monod Institute, Paris, France), and the France BioImaging infrastructure supported by the French National Research Agency (ANR-10-INSB-04, « Investments for the future »). This research was funded by Solvay, and in part by the Agence Nationale de la Recherche under the contract ANR-13-BS08-0015 (PANORAMA), ANR-12-CHEX-0011 (PULMONANO) and ANR-15-CE18-0024-01 (ICONS).

# TOC image

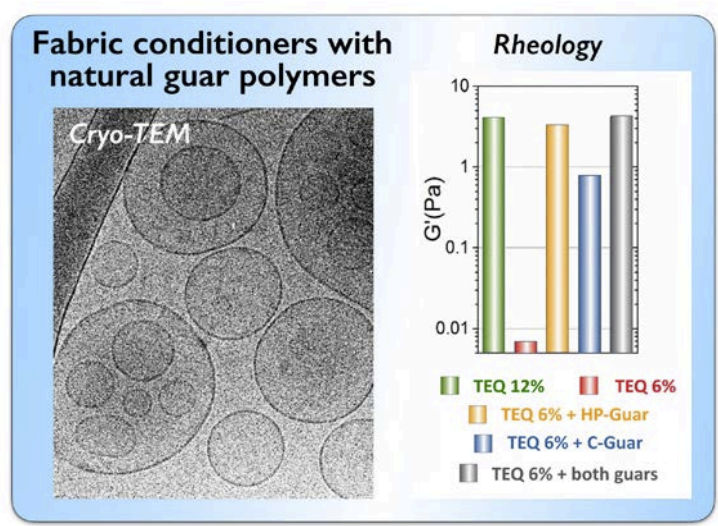